\documentclass[fleqn,twoside]{article}

\usepackage[headings]{espcrc2}

\readRCS
$Id: espcrc2.tex,v 1.2 2004/02/24 11:22:11 spepping Exp $
\usepackage{epsfig}
\usepackage{graphicx}
\newcommand{\PO}{I\!\!P}
\newcommand{\RO}{I\!\!R}
\newcommand{\xpom}{x_{\PO}}

\newcommand{\AmS}{{\protect\the\textfont2
  A\kern-.1667em\lower.5ex\hbox{M}\kern-.125emS}}

\title{Diffractive cross sections at HERA and diffractive PDFs}

\author{Laurent Schoeffel (on behalf of the H1 and ZEUS collaborations)
\address{CEA Saclay, Irfu/SPP, 
        91191 Gif-sur-Yvette Cedex, France}%
        }
       
\begin{document}

\begin{abstract}
A large collection of results for the
 diffractive  dissociation of virtual photons, $\gamma^{\star}p \to Xp$, have been obtained 
with the H1 and ZEUS detectors at HERA.
Different experimental techniques have been used,
by requiring a large rapidity gap between 
$X$ and the outgoing proton, by analysing the mass 
distribution, $M_X$, of the hadronic final state, as well as by directly 
tagging the proton. A reasonable compatibility between those techniques and between H1 and ZEUS results have been observed.
Some common fundamental features
in the measurements are also present in all data sets. 
They are detailed in this document.
Diffractive PDFs can give a good account of those features. Ideas and results are discussed in the following.
\vspace{1pc}
\end{abstract}

\maketitle

\section{Experimental diffraction at HERA}

\begin{figure}[tbp]
\begin{center}
\psfig{figure=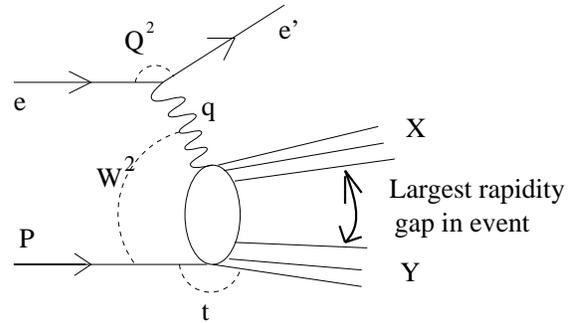,width=0.43\textwidth,angle=0}
\end{center}
\vspace{-1.cm}
\caption{Illustration of the process $ep \rightarrow eXY$. The
hadronic final state is composed of two distinct systems $X$ and
$Y$, which are separated by the largest interval 
in rapidity between final state hadrons.}
\label{difproc}
\end{figure}

One of the most important experimental result from the DESY $ep$ collider HERA
is the observation of a significant fraction of events in Deep Inelastic Scattering (DIS)
with a large rapidity gap (LRG) between the scattered proton, which remains intact,
and the rest of the final system. This fraction corresponds to about 10\% of the DIS   data
at $Q^2=10$ GeV$^2$.
In DIS, such events are not expected in such abundance, since large gaps are exponentially
suppressed due to color string formation between the proton remnant and the scattered partons.
Events
are of the type $ep \rightarrow eXp$, where the final state proton
carries more than $95$ \% of the proton beam energy. 
A photon of virtuality $Q^2$, coupled to the electron (or positron),
undergoes a strong interaction with the proton (or one of its 
low-mass excited states $Y$) to form a hadronic final state
system $X$ of mass $M_X$ separated by a LRG
from the leading proton (see Fig. \ref{difproc}). 
These events are called diffractive.
In such a reaction, $ep \rightarrow eXp$,
no net quantum number is exchanged and 
  the longitudinal momentum fraction $1-x_{\PO}$  
  is lost by the proton. Thus, the mongitudinal momentum $\xpom P$ is transfered 
to the system $X$. In addition to the standard DIS kinematic variables and $\xpom$, 
a diffractive event is also often 
characterised by the variable $\beta={x_{Bj}}/{x_{\PO}}$, which takes a simple 
interpretation in the parton model discussed in the following.

\begin{figure}[tp]
\vspace{5.3cm}
\hspace{-1.5cm}
\psfig{figure=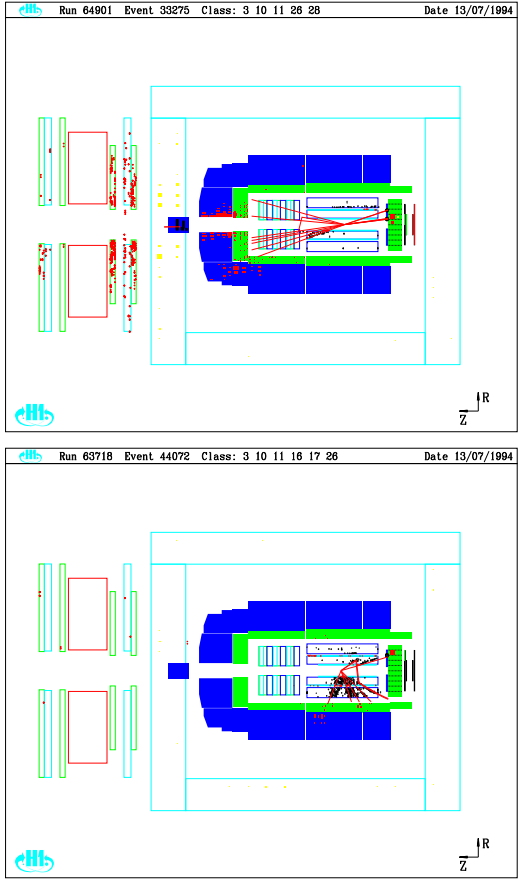,width=0.25\textwidth,angle=0}
\vspace{-2.cm}
\caption{Usual (top) and diffractive (bottom) events in the H1 experiment at HERA.
For a diffractive event, no hadronic activity is visible in the
the proton fragmentation region, as the proton remains intact 
in the diffractive process. On the contrary, for a standard DIS event,
the proton is destroyed in the reaction and the flow of hadronic clusters
is clearly visible in the proton fragmentation region (+z direction, i.e. forward part of the detector).}
\label{experimentaldiff}
\end{figure}

Experimentally,
a diffractive DIS event, $ep\rightarrow eXp$, is
presented in Fig.~\ref{experimentaldiff} (bottom). The dissociating
particle is the virtual photon emitted by the electron. The final
state consists of the scattered electron and hadrons which populate
the photon fragmentation region. The proton is scattered in the
direction of the initial beam proton with little change in 
momentum and angle. In particular, we detect no hadronic activity in the
direction of the proton flight, as the proton remains intact 
in the diffractive process. On the contrary, for a standard DIS event (Fig.~\ref{experimentaldiff} top),
the proton is destroyed in the reaction and the flow of hadronic clusters
is clearly visible in the proton fragmentation region (forward part of the detector).

\begin{figure}[tbp]
\begin{center}
\psfig{figure=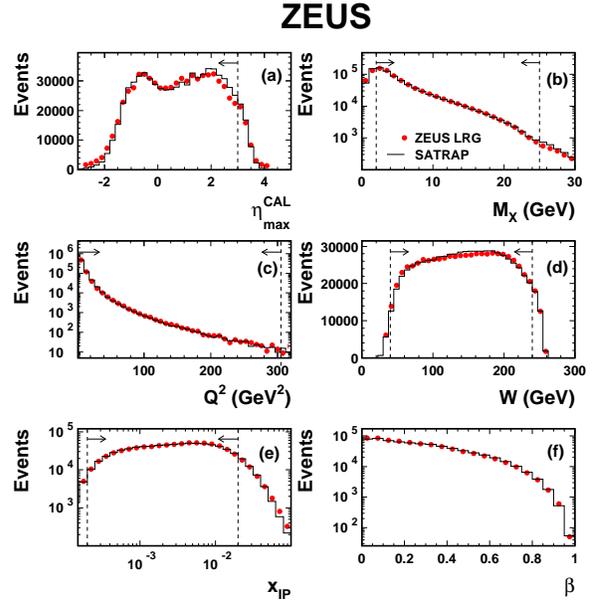,width=0.5\textwidth,angle=0}
\end{center}
\vspace{-1.3cm}
\caption{Comparison of the distributions of data (dots) to those obtained with
the Monte-Carlo (histograms) for typical variables in the LRG analysis.}
\label{datalrg}
\end{figure}

\begin{figure}[tbp]
\begin{center}
\psfig{figure=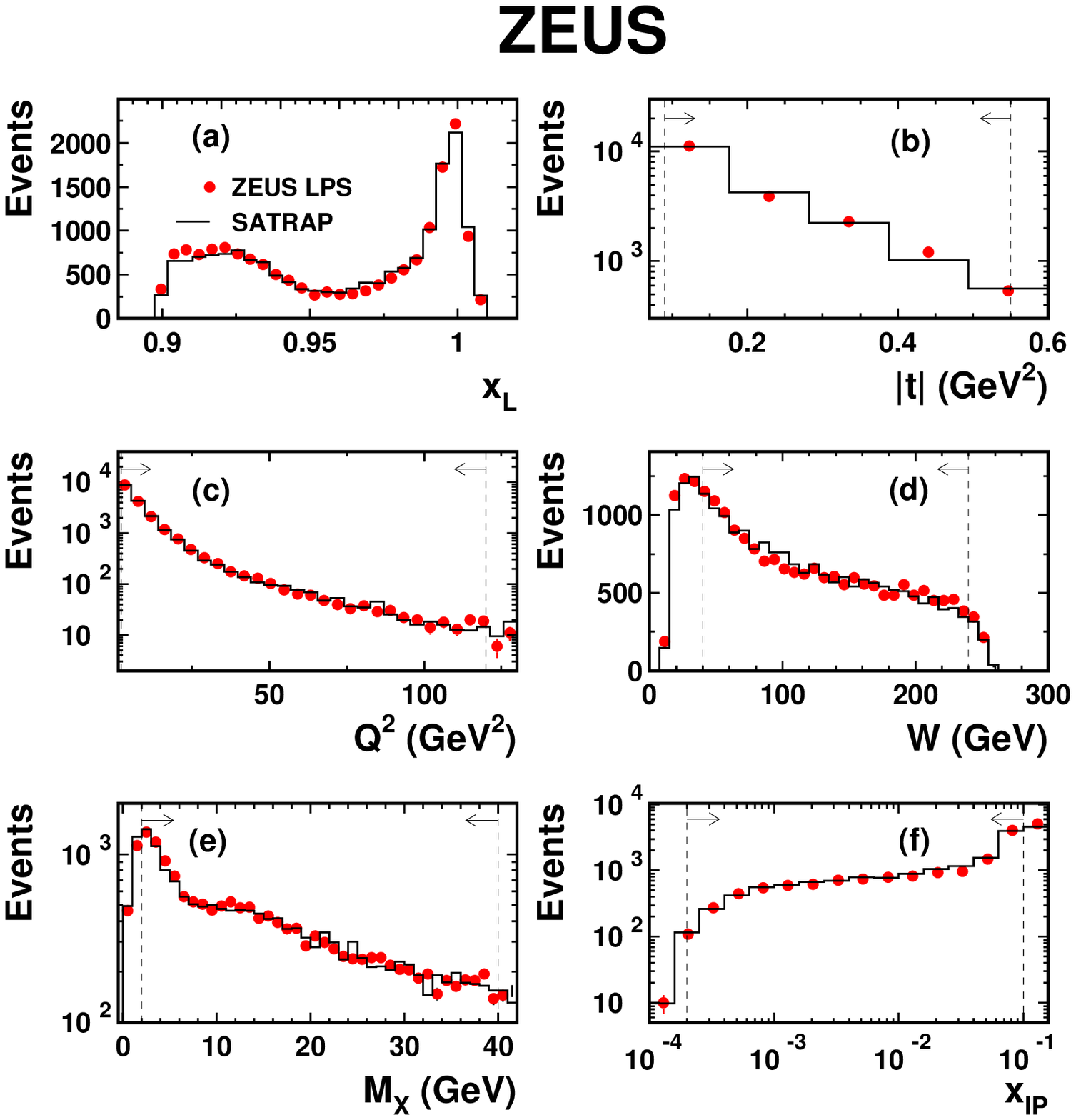,width=0.5\textwidth,angle=0}
\end{center}
\vspace{-1.3cm}
\caption{Comparison of the distributions of data (dots) to those obtained with
the Monte-Carlo (histograms) for typical variables in the LPS analysis.}
\label{datalps}
\end{figure}

\begin{figure}[tbp]
\begin{center}
\psfig{figure=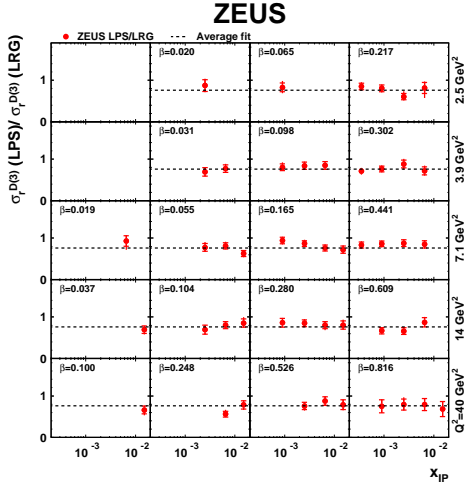,width=0.4\textwidth,angle=0}
\end{center}
\vspace{-1.3cm}
\caption{Ratio of the diffractive cross sections, as obtained with the LPS and
the LRG experimental techniques. The lines indicate the average value of the ratio,
which is about 0.86. It implies that the LRG sample contains about
24\% of proton dissociation  events, corresponding to processes like $ep \rightarrow eXY$,
where $M_Y<2.3$ GeV. This fraction is approximately the same for H1 data (of course in the same $M_Y$ range).}
\label{lpsoverlrg}
\end{figure}

\begin{figure}[tbp]
\begin{center}
\psfig{figure=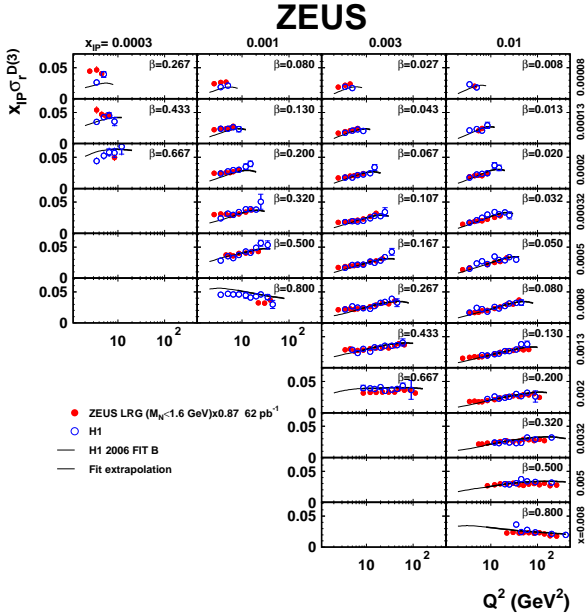,width=0.5\textwidth,angle=0}
\end{center}
\vspace{-1.3cm}
\caption{The diffractive cross sections obtained with the LRG method by the H1 and ZEUS experiments.
The ZEUS values have been rescaled (down) by a global factor of 13 \%. This value is compatible with the normalisation uncertainty
of this sample. }
\label{datah1zeus}
\end{figure}

\begin{figure}[tbp]
\begin{center}
\psfig{figure=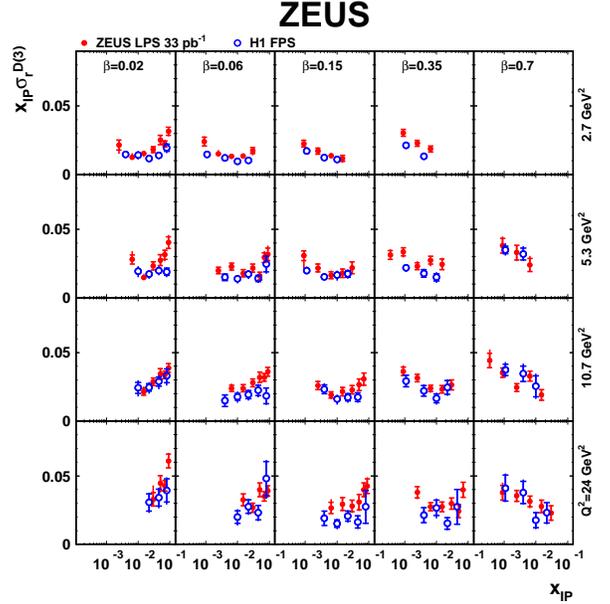,width=0.5\textwidth,angle=0}
\end{center}
\vspace{-1.3cm}
\caption{The diffractive cross section obtained with the FPS (or LPS) method by the H1 and ZEUS experiments,
where the proton is tagged. The ZEUS measurements are above H1 by a global factor of about 10\%.}
\label{datah1zeuslps}
\end{figure}

\begin{figure}[tbp]
\begin{center}
\psfig{figure=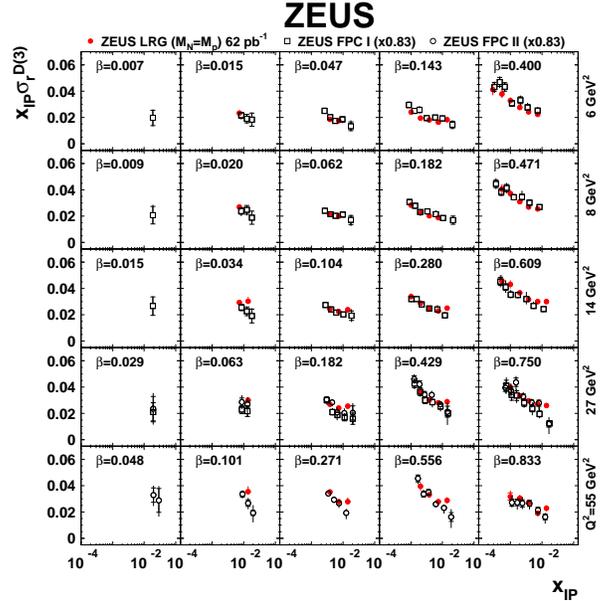,width=0.5\textwidth,angle=0}
\end{center}
\vspace{-1.3cm}
\caption{The diffractive cross sections obtained with the LRG method (full dots) compared with the
results obtained with the $M_X$ method (open symbols: FPC I and FPC II). All values are converted to
$M_Y=M_p$.}
\label{datazeuslrgmx}
\end{figure}

The experimental selection of diffractive events in DIS proceeds in two steps.  
Events are first selected based on the presence of the scattered
electron in the detector. Then, for
the diffractive selection itself, three different methods have been used at
HERA: 
\begin{enumerate} 
\item A reconstructed proton track is required in the leading (or forward) proton 
spectrometer (LPS for ZEUS or FPS for H1) with a fraction of the initial proton momentum
$x_L>0.97$. Indeed,  the cleanest selection
of diffractive events with photon dissociation is based on the
presence of a leading proton in the final state. By leading proton we
mean a proton which carries a large fraction of the initial beam
proton momentum. This is the cleanest way to select diffractive events, but
the disadvantage is a reduced kinematic coverage.
\item The hadronic system $X$ measured in the central detector is 
required to be separated by a large rapidity gap from the rest of the
hadronic final state. This is a very efficient way to select diffractive events
in a large kinematic domain, close to the standard DIS one. The prejudice is 
a large background as discussed in the following.
\item The diffractive contribution is identified as
the excess of events at small $M_X$ above the exponential fall-off of
the non-diffractive contribution with decreasing $\ln M^2_X$. The exponential fall-off, expected in
QCD, permits the subtraction of the non-diffractive
contribution and therefore the extraction of the diffractive
contribution without assuming the precise $M_X$ dependence of the
latter. This is also a very efficient way to select diffractive events
in a large kinematic domain.
\end{enumerate}
\noindent
Extensive measurements of diffractive DIS cross sections have been made by both
the ZEUS and H1 collaborations at HERA,
using different experimental techniques \cite{f2d97,f2d97b,zeus,zeusb,marta}. 
Of course, the comparison of these techniques provides a rich
source of information to get a better understanding of the experimental gains
and prejudices of those techniques.
In Fig. \ref{datalrg} and \ref{datalps}, the basis of the last ZEUS experimental analysis 
is summarised \cite{marta}. Data are compared to Monte-Carlo (MC) expectations for typical
variables. The MC is based on specific models for signal and backgrounds, 
and the good agreement with data is proof that the main ingredients of the
experimental analysis are under control: resolutions, calibrations, efficiencies...
These last sets of data (Fig. \ref{datalrg} and \ref{datalps}) \cite{marta} contain
five to seven times more statistics than in preceding publications of diffractive
cross sections, and thus opens the way to new developments in data/models comparisons.
A first relative control of the  data samples is shown in Fig. \ref{lpsoverlrg}, where the
ratio of the diffractive cross sections is displayed,
as obtained with the LPS and
the LRG experimental techniques. The mean value of the ratio of $0.86$ indicates that
the LRG sample contains about 24\% of proton-dissociation background, which is not
present in the LPS sample. This background corresponds to events like
$ep \rightarrow e X Y$, where $Y$ is a low-mass excited state of the proton (with
$M_Y < 2.3$ GeV). 
It is obviously not present in the LPS analysis which  can select specifically a proton
in the final state.   This is the main background in the LRG analysis. Due to a lack
of knowledge of this background, it causes a large normalisation uncertainty of 10  to 15 \% for the
cross sections extracted from the LRG analysis.
We can then compare the results obtained by the H1 and ZEUS experiments for diffractive
cross sections (in Fig. \ref{datah1zeus}), using the LRG method.
A good compatibility of both data sets is observed, after rescaling the ZEUS points by 
a global factor of 13\%. This factor is compatible with the normalisation uncertainty described above.
We can also compare the results obtained by the H1 and ZEUS experiments (in Fig. \ref{datah1zeuslps}),
 using the tagged proton  method (LPS for ZEUS and FPS for H1).
In this case, there is no proton dissociation background and the diffractive sample
is expected to be clean. It gives a good reference to compare both experiments. A global
normalisation difference of about 10\% can be observed in Fig. \ref{datah1zeuslps},
which can be studied with more data. It remains compatible with the normalisation
uncertainty for this tagged proton  sample.
It is interesting to note that the ZEUS measurements are globaly above the H1 data by 
 about 10\% for both techniques, tagged proton or LRG.
In Fig. \ref{datazeuslrgmx}, we compare the results using the LRG and the $M_X$ methods,
for ZEUS data alone.
Both sets are in good agreement, which shows that there is no strong bias between these
experimental techniques.
The important message at this level is not only the observation of differences
as illustrated in Fig.  \ref{datah1zeus} and   \ref{datah1zeuslps},
but the opportunity opened with the large satistics provided by the ZEUS measurements. 
Understanding discrepancies
between data sets is part of the experimental challenge of the next months. It certainly needs
 analysis of new data sets from the H1 experiment. However, already at the present
level, much can be done with existing data for the understanding of  diffraction at HERA.


\section{Diffractive PDFs  at HERA}
\begin{figure}[!]
\begin{center}
\includegraphics[width=8cm,height=8.5cm]{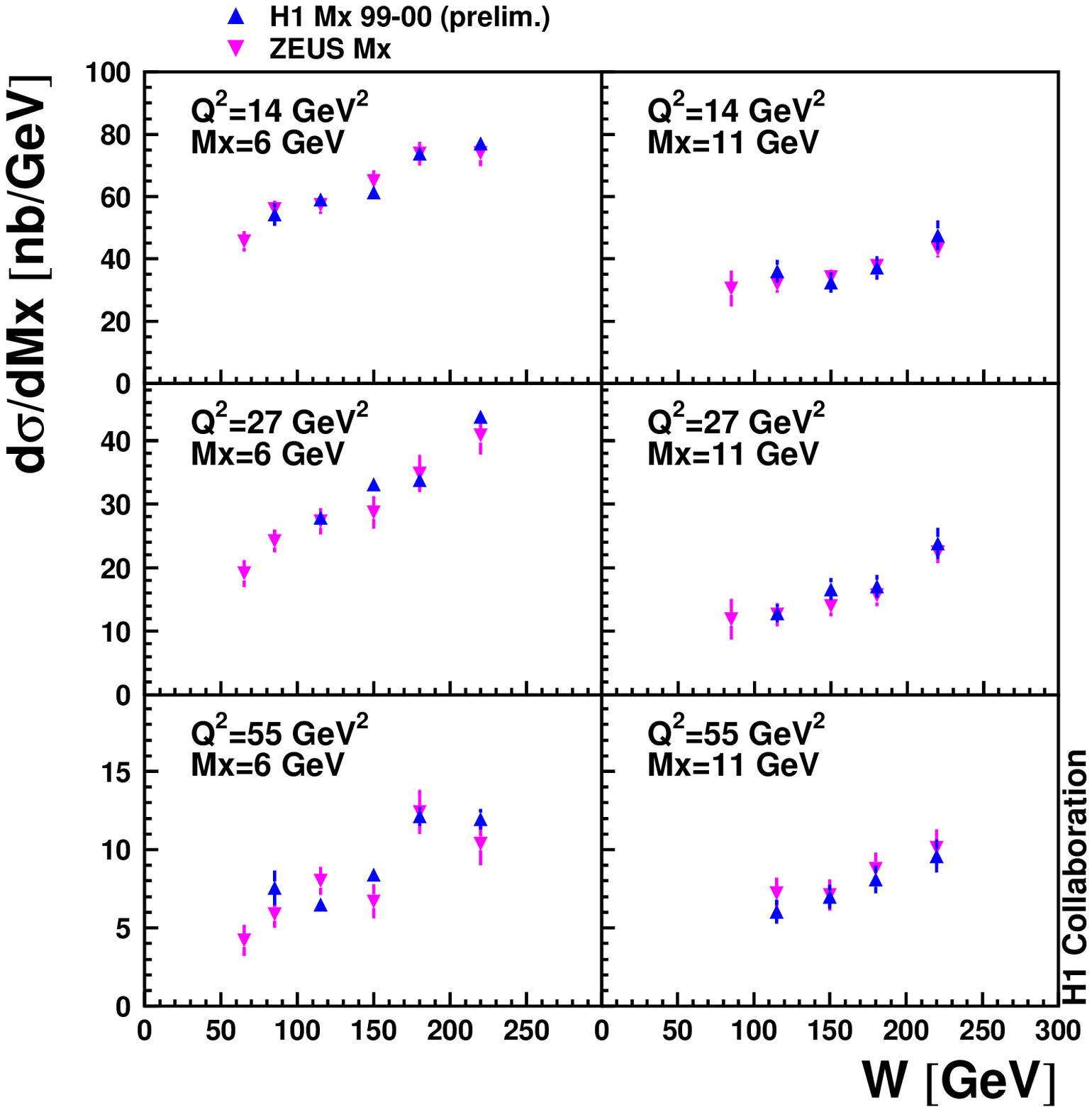}
\vspace{-1cm}
\caption{Cross sections of the diffractive process $\gamma^* p \rightarrow p' X$, 
differential in the mass of the diffractively produced hadronic system $X$ ($M_X$),
are presented as a function of the centre-of-mass energy of the $\gamma^*p$ system $W$.
Measurements at different values of the virtuality
$Q^2$ of the exchanged photon are displayed. We observe a behaviour of the form $\sim W^{ 0.6}$  for the diffractive cross section, 
compatible with the dependence expected
for a hard process. 
}
\label{figdata}
\end{center}
\vspace{-0.5cm}
\end{figure}

\begin{figure}[!]
\begin{center}
\psfig{figure=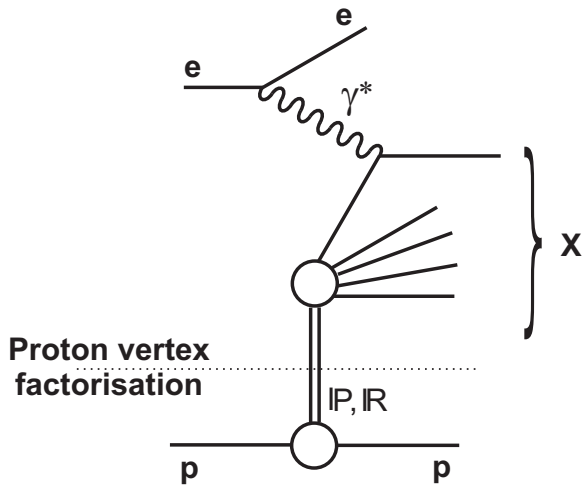,width=0.45\textwidth,angle=0}
\end{center}
\vspace{-1.3cm}
\caption{Schematic diagram of a diffractive process.
Events with a diffractive topology can be studied  in terms of
the Pomeron trajectory exchanged between the proton and the virtual photon.
}
\label{pomeron}
\end{figure}

In order to compare diffractive data with perturbative QCD models, or parton-driven models,
the first step is to show that 
the diffractive cross section
shows a hard dependence in the centre-of-mass energy $W$ of the $\gamma^*p$ system.
In Fig. \ref{figdata}, we observe a behaviour of the form $\sim W^{ 0.6}$ , 
compatible with the dependence expected
for a hard process. This  observation is obviously 
the key to allow further studies of the diffractive process in the context of
perturbative QCD.
Events with the diffractive topology can be studied  in terms of
Pomeron trajectory exchanged between the proton and the virtual photon.
In this view, these events result from a colour-singlet exchange
between the diffractively dissociated virtual photon and the proton (see Fig. \ref{pomeron}). 

A diffractive
structure function $F_2^{D(3)}$ 
can then be defined as a sum of two factorized 
contributions, corresponding to a Pomeron and secondary Reggeon trajectories: \\
$
F_2^{D(3)}(Q^2,\beta,x_{\PO})=
f_{\PO / p} (x_{\PO}) F_2^{D(\PO)} (Q^2,\beta) \
+ f_{\RO / p} (x_{\PO}) F_2^{D(\RO)} (Q^2,\beta)
$,
where $f_{\PO / p} (x_{\PO})$ is the Pomeron flux. It depends only on $\xpom$,
once integrated over $t$, and
$F_2^{D(\PO)}$ can be interpreted as the Pomeron structure function,
depending on $\beta$ and $Q^2$.
The other function,
$F_2^{D(\RO)}$, is an effective Reggeon structure function
taking into account various secondary Regge contributions which can not be 
separated.
The Pomeron and Reggeon fluxes are assumed to follow a Regge behaviour with  
linear
trajectories $\alpha_{\PO,\RO}(t)=\alpha_{\PO,\RO}(0)+\alpha^{'}_{\PO,\RO} t$, 
such that
\begin{equation}
f_{{\PO} / p,{\RO} / p} (x_{\PO})= \int^{t_{min}}_{t_{cut}} 
\frac{e^{B_{{\PO},{\RO}}t}}
{x_{\PO}^{2 \alpha_{{\PO},{\RO}}(t) -1}} {\rm d} t ,
\label{flux}
\end{equation}
where $|t_{min}|$ is the minimum kinematically allowed value of $|t|$ and
$t_{cut}=-1$ GeV$^2$ is the limit of the measurement. 
We take
$\alpha^{'}_{\PO}=0.06$ GeV$^{-2}$, 
$\alpha^{'}_{\RO}=0.30$ GeV$^{-2}$,
$B_{\PO}=5.5$ GeV$^{-2}$ and $B_{\RO}=1.6$ GeV$^{-2}$. 
The Pomeron
intercept $\alpha_{\PO}(0)$ is left as a free parameter in the QCD fit
and $\alpha_{\RO}(0)$ is fixed to $0.50$.

\begin{figure}[!]
\begin{center}
\psfig{figure=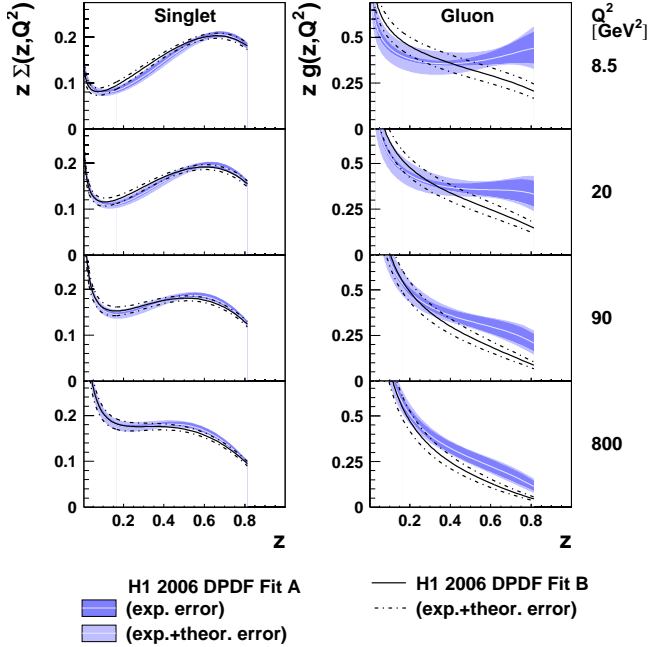,width=0.5\textwidth,angle=0}
\end{center}
\vspace{-1.3cm}
\caption{Singlet and gluon distributions 
of the Pomeron (DPDFs) as a function of $z \equiv \beta$, the fractional momentum of the
Pomeron carried by the struck parton (see text), obtained by a QCD fit to the H1 
diffractive cross sections.}
\label{dpdfsh1}
\end{figure}

\begin{figure}[!]

\centerline{\includegraphics[width=0.9\columnwidth]{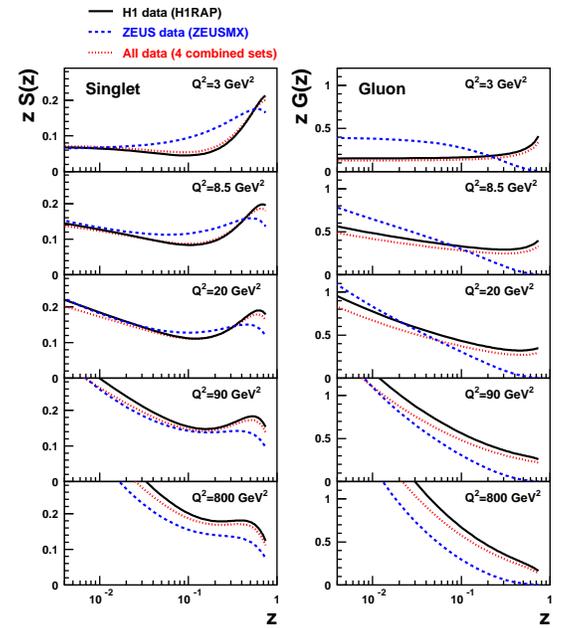}}
\vspace{-1.cm}
\caption{Singlet and gluon DPDFs as a function of $z \equiv \beta$, where the results of fitting H1 or ZEUS
data are compared. 
The ZEUS data considered here \cite{zeus} are derived using 
the $M_X$ method.
A global fit of all published data is also presented. Note that
the last ZEUS data set \cite{marta} is not used for this plot.}
\label{dpdfsperso}
\end{figure}

\begin{figure}[!]
\begin{center}
\psfig{figure=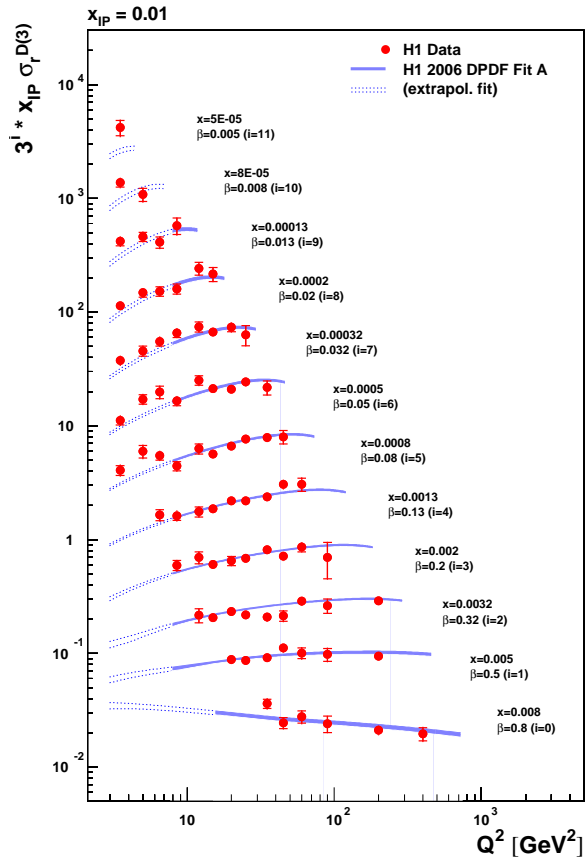,width=0.45\textwidth,angle=0}
\end{center}
\vspace{-0.5cm}
\caption{Scaling violations for H1 diffractive cross sections for one value of $\xpom$ ($\xpom=0.01$)
and a large range of  $\beta$ values, from low  ($<0.01$) to large values ($> 0.5$).}
\label{scalingviol}
\end{figure}

The next step is then to model the Pomeron structure function $F_2^{D(\PO)}$
\cite{f2d97,lolo1,lolo2,lolo3}.
Among the most popular models, the one based on a pointlike structure of
the Pomeron has been studied extensively 
using a non-perturbative input supplemented by a perturbative QCD evolution equations
\cite{lolo1,lolo2,lolo3}.
In this formulation, it is assumed that the exchanged object, the Pomeron, 
is a colour-singlet quasi-particle whose structure is probed in the
DIS process. 
As for standard DIS,   diffractive parton distributions  
related to the Pomeron can be derived from QCD fits to diffractive cross sections.
The procedure is standard: we assign parton distribution functions to the Pomeron
parametrised in terms of non-perturbative input
distributions at some low scale $Q_0^2$. The quark flavour singlet distribution
($z{ {S}}(z,Q^2)=u+\bar{u}+d+\bar{d}+s+\bar{s}$)
and the gluon distribution ($z{\it {G}}(z,Q^2)$) are parametrised 
at this initial scale
$Q^2_0$, where $z=x_{i/I\!\!P}$ is the fractional momentum of the Pomeron carried by
the struck parton. Functions $z{\it{S}}$ and $z{\it{G}}$  
are evolved to higher $Q^2$ using the
next-to-leading order DGLAP evolution equations.
For the structure of the sub-leading Reggeon trajectory,
the pion structure function 
\cite{owens} is assumed
with a free global normalization to be determined by the data. 
Diffractive PDFs (DPDFs) extracted from H1 and ZEUS data are shown in Fig. \ref{dpdfsh1} and \ref{dpdfsperso}
\cite{f2d97,lolo1,lolo2,lolo3}. 
We observe that some differences in the data are reflected in the DPDFs, but some
basic features are common for all data sets and the resulting DPDFs. Firstly, the gluon density
is larger than the sea quark density, which means that the major fraction of the momentum
(about 70\%) is carried by the gluon for a typical value of $Q^2=10$ GeV$^2$. Secondly, we observe 
that the gluon density is quite large at large $\beta$, with a large uncertainty, which means that
we expect positive scaling violations still at large values of $\beta$. This is shown in Fig. \ref{scalingviol}.
We note that even at large values of $\beta \sim 0.5$, the scaling violations are still positive,
as discussed above. The strength of the DPDFs approach is to give a natural interpretation of this
basic observation and to describe properly the $Q^2$ evolution of the cross sections.
Other approaches are also well designed to describe all features of the data \cite{plb}, but this
is another story.
The near future of the study of DPFDs is to combine all existing data and check their
compatibility with respect to the QCD fit technique. If this is verified, a new global analysis
can be followed to get the most complete understanding of DPDFs \cite{lolo1}.


\section{Diffractive PDFs and the LHC}

\begin{figure}[!]
\begin{center}
\epsfig{file=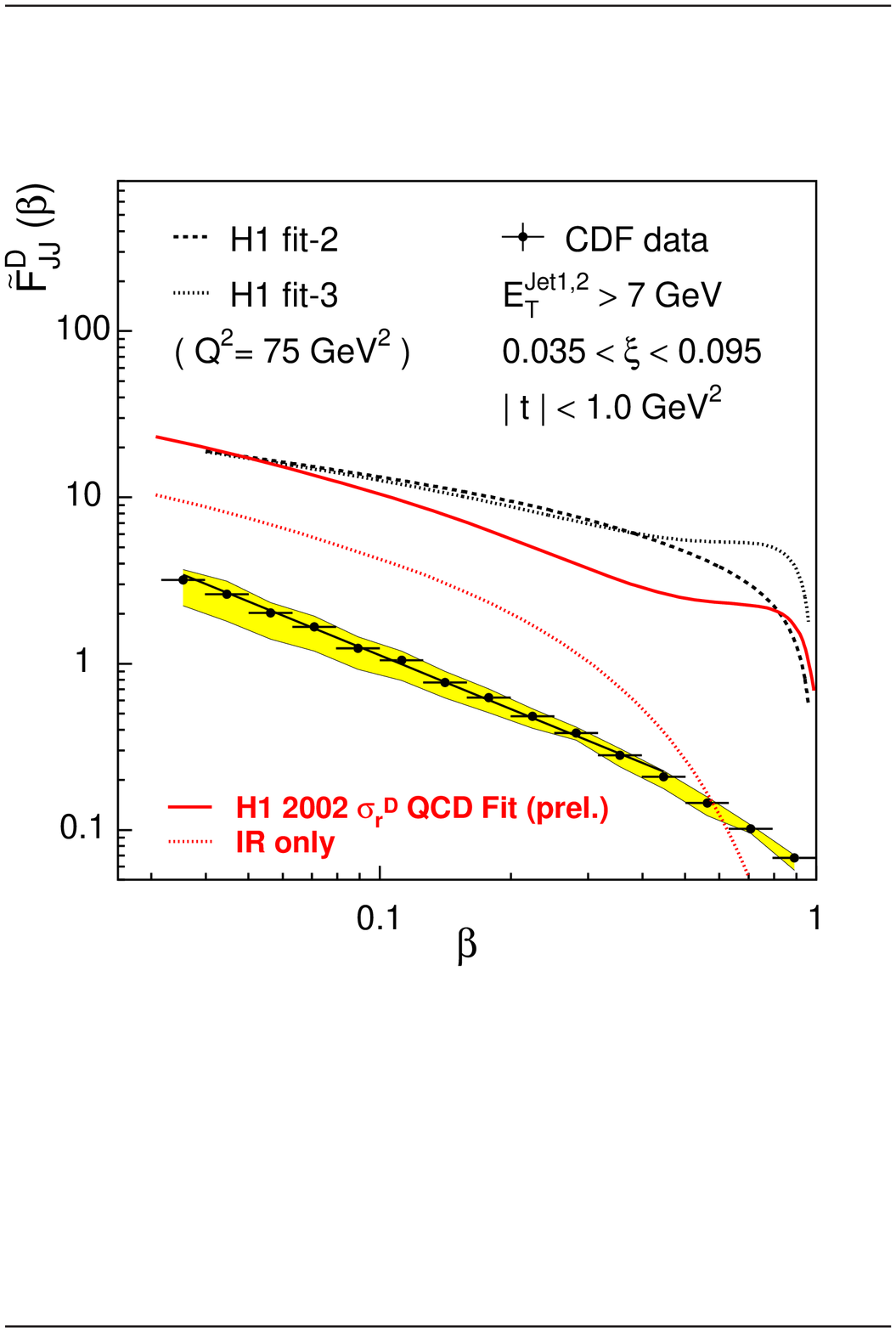,width=7cm,clip=true}
\vspace{6.cm}
\caption{Comparison between the CDF measurement of diffractive structure
function (black points) with the H1 diffractive PDFs.}
\label{cdfh}
\end{center}
\end{figure} 


Note that diffractive distributions are process-independent
functions.  They appear not only in inclusive diffraction but also in
other processes where diffractive hard-scattering factorisation holds.  
The cross section of such a process can be
evaluated as the convolution of the relevant parton-level
cross section with the DPDFs.
For instance, the cross section
for charm production in diffractive DIS can be calculated at leading order
in $\alpha_s$ from the $\gamma^* g \rightarrow c \bar c$ cross section and
the diffractive gluon distribution.  An analogous statement holds for jet
production in diffractive DIS. Both processes have been analysed at
next-to-leading order in $\alpha_s$ and are found to be consistent with
the factorisation theorem \cite{collins}.
A natural question to ask is whether one can use the DPDFs
extracted at HERA to describe hard diffractive processes such as the
production of jets, heavy quarks or weak gauge bosons in $p\bar{p}$
collisions at the Tevatron.  Fig.~\ref{cdfh} shows results on
diffractive dijet production from the CDF collaboration
compared to the expectations based on the 
DPDFs from HERA \cite{cdf}.  The discrepancy is spectacular:
the fraction of diffractive dijet events at CDF is a factor 3 to 10
smaller than would be expected on the basis of the HERA data. The same
type of discrepancy is consistently observed in all hard diffractive
processes in $p\bar{p}$ events.  In
general, while at HERA hard diffraction contributes a fraction of order
10\% to the total cross section, it contributes only about 1\% at the
Tevatron.
This observation of QCD-factorisation breaking in hadron-hadron
scattering can be interpreted as a survival gap probability or a soft color interaction
which needs
to be considered in such reactions.
In fact, from a fundamental point of view, diffractive hard-scattering factorization does not apply to
hadron-hadron collisions.
Attempts to establish corresponding factorization theorems fail,
 because of interactions between spectator partons of the colliding
   hadrons.  The contribution of these interactions to the cross section
   does not decrease with the hard scale.  Since they are not associated
   with the hard-scattering subprocess, we no
   longer have factorization into a parton-level cross section and the
   parton densities of one of the colliding hadrons. These
interactions are generally soft, and we have at present to rely on
phenomenological models to quantify their effects \cite{cdf}. 
The yield of diffractive events in hadron-hadron collisions is then lowered
precisely because of these soft interactions between spectator partons
(often referred to as reinteractions or multiple scatterings).  
They can produce additional final-state particles which fill the would-be
rapidity gap (hence the often-used term rapidity gap survival).  When
such additional particles are produced, a very fast proton can no longer
appear in the final state because of energy conservation.  Diffractive
factorization breaking is thus intimately related to multiple scattering
in hadron-hadron collisions. Understanding and describing this
phenomenon is a challenge in the high-energy regime that will be reached
at the LHC \cite{afp}.
We can also remark simply that
the collision partners, in $pp$ or $p\bar{p}$ reactions, are both
composite systems of large transverse size, and it is not too
surprising that multiple interactions between their constituents can
be substantial.  In contrast, the virtual photon in $\gamma^* p$
collisions has small transverse size, which disfavors multiple
interactions and enables diffractive factorization to hold.  According
to our discussion, we may expect that for
decreasing virtuality $Q^2$ the photon behaves more and more like a
hadron, and diffractive factorization may again be broken.

\section{Conclusions}

We have presented and discussed the most recent results on
inclusive diffraction from the HERA experiments.
A large collection of data sets and diffractive cross sections 
are published, which present common fundamental
features in all cases. 
The different experimental techniques, for both H1 and ZEUS
experiments, provide compatible results, with still some global normalisation
differences of about 10\%.
DPDFs give a good account of the main features of the diffractive data.
There is still much to do on the experimental side
with large statistics analyses, in order to obtain a better understanding
of  data and backgrounds. This is an essential task for the next
months with the purpose to understand and reduce the normalisation
uncertainties of diffractive measurements at HERA. This will make the
combination of cross sections between the two experiments much easier,
with a common message from HERA on inclusive diffraction.


\end{document}